\documentclass[journal]{IEEEtran}
\usepackage{cite}

\usepackage{graphicx}
\usepackage{amsmath}
\usepackage{algorithm,algorithmic}

\usepackage{array}

  \usepackage[caption=false,font=footnotesize]{subfig}
\usepackage{url}

\usepackage{makecell}
\usepackage{xcolor}

\renewcommand{\baselinestretch}{0.975}

\begin{document}
%
\title{Automated Verification of Stand-alone Solar Photovoltaic Systems}
%
%
%

\author{Alessandro~Trindade and Lucas~Cordeiro
\thanks{A. Trindade is with the Department of Electricity, Federal University of Amazonas, Manaus, Brazil, e-mail: alessandrotrindade@ufam.edu.br.}
\thanks{L. Cordeiro is with School of Computer Science, The University of Manchester, UK, e-mail: lucas.cordeiro@manchester.ac.uk.}
\thanks{The authors would like to thank Newton Fund (ref. 261881580) and FAPEAM (Amazonas State Foundation for Research Support, call 009/2017), for the financial support.}
\thanks{}}

%
%

\markboth{}
{Trindade \MakeLowercase{\textit{et al.}}: AUTOMATED VERIFICATION OF STAND-ALONE SOLAR PHOTOVOLTAIC SYSTEMS}
%



\maketitle

\begin{abstract}
With declining costs and increasing performance, the deployment of renewable energy systems is growing faster. Particular attention is given to stand-alone solar photovoltaic systems in rural areas or where grid extension is unfeasible. Tools to evaluate electrification projects are available, but they are based on simulations that do not cover all aspects of the design space. Automated verification using model checking has proven to be an effective technique to program verification. This paper marks the first application of software model checking to formally verify the design of a stand-alone solar photovoltaic system including solar panel, charge controller, battery, inverter, and electric load. Case studies, from real photovoltaic systems deployed in five different sites, ranging from 700W to 1,200W, were used to evaluate this proposed approach and to compare that with specialized simulation tools. Data from practical applications show the effectiveness of our approach, where specific conditions that lead to failures in a photovoltaic solar system are only detected by our automated verification method.
\end{abstract}


%
\IEEEpeerreviewmaketitle

\section{Introduction}
%
%
%
%
\IEEEPARstart{A}{ccording} to Coelho et al.~\cite{Coelho}, there are presently 1.3 billion people with no access to electricity worldwide. 
%
Only a niche market a few years ago, solar photovoltaic (PV) systems are now becoming a mainstream electricity provider. 
There was an increase of approximately 50\% from 2016 to 2017 in terms of new installations of PV all over the world \cite{EPIA}. That scenario brings the need for 
%
%
%
design validation -- ensuring the correctness of the design at earliest stages -- which is a major challenge in any responsible system development process, and the activities intended for its solution occupy an ever increasing portions of the  development cycle cost and time budgets \cite{ClarkeHV18}. 

In order to model, simulate, or evaluate a PV system, there are a myriad of specialized tools available in the market such as RETScreen, HOMER, PVWatts, SAM, and Hybrid2 \cite{Pradhan,Swarnkar,NRELDobos,NRELBlair,Mills}; and even general purpose simulation tools such as PSpice, Saber, or MATLAB/Simulink package \cite{Gow1999,Benatiallah2017}.
%
%
%
%
%
However, those tools are based on running experiments in simulation models. Simulation has the advantage of being cheap (if compared to test in real systems) and can be employed before the system design is concluded but it has the drawback of an incomplete coverage since the verification of all possible combinations and potential failures of a system is rarely possible or even unfeasible~\cite{ClarkeHV18} to be achieved in practice.

Formal methods based on model checking offer a great potential to obtain a more effective and faster verification in the design process~\cite{ClarkeHV18}.  
Any kind of system can be specified as computer programs using mathematical logic, which constitutes the intended (correct) behavior; then, one can try to give a formal proof or otherwise establish that the program meets its specification. User or project requirements can be added during the creation of the formal model to be verified. 
%
%
Model checking algorithms can then verify the system model by systematically exploring all its states to check whether the requirements are met by the given system.
%
%
In this study, a mathematical model of each component of a stand-alone PV system, as panel solar, charge controller, batteries, inverter, and electrical load are created. The behavior of each system component can be analyzed and observed with the support of those formal models, as a joint operation of the components, which in this case represents the operation of the solar PV system itself. A key benefit to this approach is that it helps in the detection of flaws in the design phase of system development, thereby considerably improving system reliability~\cite{Akram2018}.

Related to solar PV systems, the project requirements, as battery autonomy and power demand, besides weather conditions, as solar irradiance and ambient temperature, are translated as part of the computer program and automatically verified during the formal process. The model checking tool reports in which conditions a system does not meet the user requirements or whether it will fail due to weather conditions, which aids to improve the project itself.  
The implementation of the proposed tool is carried out by means of the efficient SMT-based bounded model checker (ESBMC)~\cite{esbmc2018}, which allows one to incrementally verify a PV system as an imperative program using a fragment of decidable first-order theories~\cite{DBLP:books/daglib/0019162}.

In prior studies, the evaluation of PV systems w.r.t. user requirements were performed by software simulation tools using MATLAB/Simulink~\cite{Benatiallah2017, Samrat2014, Natsheh2012}, or HOMER Pro \cite{Lamnadi2017}. Some related studies were carried out toward the formal modeling of power smart grids~\cite{Akram2018} and to maximize the power point of solar panels~\cite{Driouich2017}; however, those studies do not perform automated formal verification and they restrict themselves to solar panels or smart grids. 
In addition, recent research that applies formal verification to solar energy, has attempted to formalize and implement a formal study of large population of PV panels, where the focus has been on the modeling of the dynamics of PV panels and their interaction with the grid, without batteries~\cite{Abate2017}; or to model a PV system in Modelica, to verify the maximum power point of solar panels with the use of Jmodelica tool~\cite{Driouich2018}. Both studies restrict to PV panels, and do not include batteries, inverters, and charge controllers.
  
Given the current knowledge in formal verification, this is the first study to apply a formal verification technique to formally check the design of a stand-alone solar PV system. In summary, this paper makes three original contributions. Firstly, we describe a modular modeling of each component of a PV system by means of mathematical models that can be encoded into fragments of first-order theories supported by software model checkers. Secondly, we propose an automated verification method that formally checks the design of a given PV system using incremental model checking based on Satisfiability Module Theories (SMT). Thirdly, experimental results show that this proposed approach can find subtle design errors in PV systems, which are not easily detected by other state-of-the-art approaches based on simulation. 

%
 
\textit{Outline}. Section~\ref{sec:SolarPhotovoltaicSystem} gives the background about solar PV systems, design and validation of PV systems, and the mathematical modeling. Section~\ref{sec:AutomatedVerification} presents the automated verification technique. The methodology is presented in section~\ref{sec:Methodology}. Section~\ref{sec:results} is devoted to the results. Section~\ref{sec:Conclusions} presents the conclusion and describes future work.

\section{Solar Photovoltaic System }
\label{sec:SolarPhotovoltaicSystem}


PV systems are classified into three distinct types~\cite{Mohanty}: (1) stand-alone systems, where the energy is generated and consumed in the same place and which do not interact with the main grid; (2) grid-connected systems; and (3) solar PV hybrid system. 
%
Specifically for the energy needed for remote rural areas of developing countries or places where the grid extension is not possible or even feasible, 
the most suitable configuration is the regulated stand-alone system with battery and AC load; this configuration is the focus of this study.

\subsection{Design and Simulation of Solar PV systems}
%
In order to address different aspects of the PV system design, there are various software tools available in the literature~\cite{Rajanna,Rawat}.
The capabilities of those tools range from simple solar resource and energy production estimation, 
 to complex financial analysis and project optimization. 
%
Here we evaluated the most popular ones: PVWatts, SAM, HOMER, RETScreen, and Hybrid2~\cite{Pradhan,Swarnkar,NRELDobos,NRELBlair,Mills}.

Table~\ref{table:softwares} summarizes the off-the-shelf tools employed here, where only Hybrid2 does not have technical support; HOMER and Hybrid2 perform off-grid system or battery backup analysis. 
Additionally, HOMER and RETScreen include economical analysis or even optimization-sensitive analysis. RETScreen and HOMER have a free web-based version, but they have limited resources since they do not allow us to save the PV projects or even upload data from manufacturers. However, commercial version of those tools, called RETScreen Expert and HOMER Pro, are available only for Microsoft Windows and the annual subscription typically range from US\$504.00 to US\$657.00.

\begin{table}[!t]
\renewcommand{\arraystretch}{1.3}
\caption{Comparative coverage of reference software}
\label{table:softwares}
\centering
\begin{tabular}{c | c | c | c | c | c}
\hline
\hline
Characteristic  & \rotatebox{90}{PVWatts} & \rotatebox{90}{SAM} & \rotatebox{90}{HOMER} & \rotatebox{90}{RETScreen} & \rotatebox{90}{Hybrid2}\\
\hline
\hline
Support & X & X & X & X &  \\
\hline
Off-grid systems &   &   & X & X & X\\
\hline
Hybrid systems &  &  & X & X & X\\
\hline
Photovoltaics & X & X & X & X & X\\
\hline
Batteries &  &  & X &  & X\\
\hline
\makecell{Main technical (T) \\ or economical(E)} & T & T & E & E & T \\
\hline
Optimization &  &  & X & X &  \\
\hline
Sensitive analysis &  &  & X & X & \\
\hline
\hline
\end{tabular}
\end{table}

%
In this study, only HOMER remains for a comparative evaluation with our proposed verification approach. 
Thus, the main challenge here is to demonstrate the application of software model checking to formally verify a stand-alone PV solution, thus proving that this approach is more effective and complete than other state-of-the-art tool such as HOMER Pro. The comparative evaluation between HOMER Pro and our approach is presented in Section~\ref{sec:results_indeed}.

\subsection{Component models for a stand-alone PV system }

\begin{figure}[h]
\includegraphics[width=0.48\textwidth]{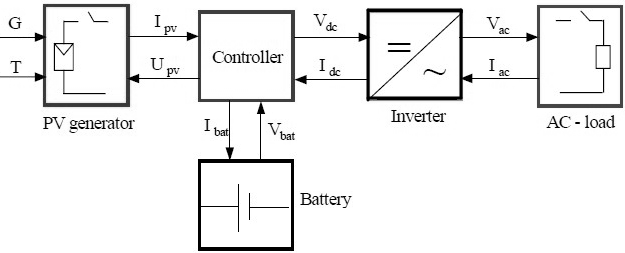}
\centering
\caption{Block diagram for a typical stand-alone PV system~\cite{Hansen}.}
\label{fig:blockdiagram}
\end{figure}

The mathematical modeling of the PV system is based on modular blocks, as illustrated in Fig.\ref{fig:blockdiagram}. It identifies the PV generator, batteries, charge controller, inverter, and AC load. 

The PV generator, which can be a panel or an array, is a semiconductor device that can convert solar energy into DC electricity. In Fig.\ref{fig:blockdiagram}, there are two variables that depend on the site where the system is deployed and the weather (i.e., solar irradiance $G$ and temperature $T$). For night hours or rainy days, we need to hold batteries, where power can be stored and used. The use of battery as a storage form implies the presence of a charge controller~\cite{Hansen,Mellit}. The PV arrays produce DC and therefore when the PV system contains an AC load, a DC/AC conversion is required. That converter is called of inverter; and the AC load dictates the behavior of AC electrical load from the house that will be fed by the system.
%
%


\subsection{PV generator model}
\label{sec:PVmodel}

The performance of PV systems 
is usually studied using an equivalent circuit model~\cite{Yatimi,Saloux,Mellit}, which consists of a current source with one or two diodes connected in parallel, and up to two resistors, one connected in parallel and the other one in series, to take into account energy losses in this model~\cite{Cubas}. 
\begin{figure}[h]
\includegraphics[width=0.20\textwidth]{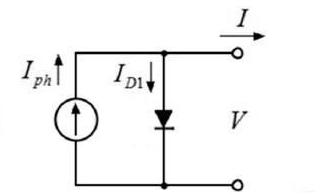}
\centering
\caption{1-diode equivalent PV cell/panel circuit model~\cite{Cubas}.}
\label{fig:equivckt}
\end{figure}
 
The 1-diode model, illustrated in Fig. \ref{fig:equivckt}, whose equation relates the output current, $I$, to the output voltage, $V$, is described by Eq.~\eqref{eq:1diodemodel}:
\begin{equation}
\label{eq:1diodemodel}
I = I_{ph}-I_{D1}=I_{ph}-I_{0}\left[ exp \left( \dfrac{V}{NaV_{T}} \right)  \right], 
\end{equation}

\noindent where $I_{ph}$ is the photocurrent delivered by the constant current source; $I_{0}$ is the reverse saturation current corresponding to the diode; $N$ is the number of series-connected cells ($N=1$ in a single cell configuration); $a$ is the ideality factor (or quality factor) that takes into account the deviation of the diodes from the Shockley diffusion theory ($a=1$ for ideal diodes and between $1$ and $2$ for real diodes); $V_{T}$ is the thermal voltage ($ V_{T}=k_{B}T/q $); $ k_{B} $ is the Boltzmann constant ($ 1.3806503\times10^{-23}J $); $T$ the temperature of the p-n junction (or cell temperature) in Kelvin; $q$ is absolute value of the electron's charge ($ -1.60217646\times10^{-19}C $).

The simplified model of 1-diode has demonstrated that it has a small error rate, between 0.03\% and 4.68\% from selected PV panels tested~\cite{Saloux}. In addition, this mathematical modeling has the advantage of being an explicit model, which does not use iterative numerical calculation, which is time-consuming to computing~\cite{Cubas}. 
 %
%
Eq.~\eqref{eq:1diodemodel} is used to express currents and voltages at each key point of the characteristic curve from a PV cell \cite{Villalva}.
The voltage and the current at the maximum power point tracking (MPPT), can be described by Equations \eqref{eq:Vmfinal}, \eqref{eq:Imfinal}, and \eqref{eq:Pm}  as follows~\cite{Saloux}: 
\begin{equation}
\label{eq:Vmfinal}
V_{m}=\dfrac{aNk_{B}T}{q} ln \left( \dfrac{aNk_{B}T}{qI_{0}} \dfrac{I_{sc}}{V_{oc}}  \right). 
\end{equation}

\begin{equation}
\label{eq:Imfinal}
I_{m} = I_{ph} + I_{0} - \dfrac{aNk_{B}T}{q} \left( \dfrac{I_{sc}}{V_{oc}} \right).  
\end{equation}

\begin{equation}
\label{eq:Pm}
P_{m} = V_{m} I_{m}.
\end{equation}

%
However, the photocurrent delivered by the constant current source ($I_{ph}$) or even the reverse saturation current ($ I_{0} $) are not given by PV manufacturers. Therefore, Eq.~\eqref{eq:Iph} is used to calculate the photocurrent as function of irradiance and temperature~\cite{Villalva}:
\begin{equation}
\label{eq:Iph}
I_{ph}=\dfrac{G}{G_{ref}} \left[ I_{ph,ref} + \mu_{I} \left( T-T_{ref} \right)    \right], 
\end{equation}

\noindent where the reference state (STC) of the cell is given by the solar irradiance $ G_{ref}=1000 W/m^{2} $ and the temperature $ T_{ref}=298.15 K (=25^{o}C) $; $ \mu_{I} $ is the short-circuit current temperature coefficient ($A/K$) 
(provided by PV manufacturers). $ I_{ph,ref} $ can be approximated to the reference short-circuit current \cite{Jakhrani} that is provided by PV manufacturers ($ I_{sc,ref} $).
The cell temperature ($ T $) is described by Eq.~\ref{eq:Tcell}~\cite{Ross}:
\begin{equation}
\label{eq:Tcell}
T = T_{air} + \dfrac{NOCT-20}{800}G,
\end{equation}

\noindent where $ T_{air} $ is the ambient temperature, $NOCT$ is the nominal operating cell temperature (in $^{o}$C) that is found at the PV manufacturer's data-sheet \cite{Ross}, and $G$ is the solar irradiance ($ W/m^{2} $) of the place where the PV system is deployed.

Furthermore, Eq.~\eqref{eq:I0} permits the saturation current ($ I_{0} $) to be expressed as a function of the cell temperature as~\cite{Villalva} 

\begin{equation}
\label{eq:I0}
I_{0} = \dfrac{I_{sc,ref} + \mu_{I}(T - T_{ref})}{exp \left[ \dfrac{q(V_{oc,ref} + \mu_{V} (T - T_{ref}))}{aNk_{B}T}    \right] -1},
\end{equation}

\noindent where $ V_{oc,ref} $ is the reference open-circuit voltage and $ \mu_{V} $ is an open-circuit voltage temperature coefficient ($ V/K $).

Using the maximum power point current (cf. Eq.~\eqref{eq:Pm}) and the saturation current in the reference temperature given by Eq.~\eqref{eq:I0}, the diode ideality factor is determined by Eq.~\eqref{eq:a}:
\begin{equation}
\label{eq:a}
a = \dfrac{q(V_{m,ref}-V_{oc,ref})}{Nk_{B}T} \dfrac{1}{ln \left( 1 - \dfrac{I_{m,ref}}{I_{sc,ref}}  \right) },
\end{equation}

\noindent where $V_{mref}$, $V_{oc,ref}$, $I_{m,ref}$, and $I_{sc,ref}$ are key cell values obtained under both actual cell temperature and solar irradiance conditions, usually provided by manufacturers; the PV generator model is now completely determined.
%
%
%
%
%
%

In addition to the model verification performed by the proposed technique, there is the prior stage of PV system sizing check, based on manufacturer's data and information from the sizing and the site; this stage ensures that the system meets its specification, thereby considering the standard project steps~\cite{Pinho}.
Firstly, we need to correct the energy consumption estimated to the load ($E_{consumption}$), which is carried out by Eq.~\eqref{eq:Ecorrected}~\cite{Pinho}, where the efficiency of batteries ($\eta_{b}$), controller ($\eta_{c}$), and the inverter ($\eta_{i}$) are considered as
\begin{equation}
\label{eq:Ecorrected}
E_{corrected} = \dfrac{E_{consumption}}{ \eta_{b} \eta_{c} \eta_{i} }.
\end{equation}

The total minimum number of needed solar panels ($N_{TPmin}$) is computed by Eq.~\eqref{eq:NTPmin} and the check is performed using Eq.~\eqref{eq:NTP}, where the sized number of panels ($ N_{TP} $) must be greater than the result from Eq.~\eqref{eq:NTPmin}.
\begin{equation}
\label{eq:NTPmin}
N_{TPmin} = \dfrac{E_{corrected}}{E_{p}}.
\end{equation}

\begin{equation}
\label{eq:NTP}
N_{TP} \geq N_{TPmin}.
\end{equation}

Particularly, the total number of panels in series ($N_{PSmin}$) and parallel ($N_{PPmin}$) are given by~\eqref{eq:NPSmin} and \eqref{eq:NPPmin}, respectively. With the check performed by (\ref{eq:NPS}) and (\ref{eq:NPP}), $ V_{system} $ is the DC voltage of the bus, normally $12$, $24$ or $48$ V.
\begin{equation}
\label{eq:NPSmin}
N_{PSmin} = \dfrac{V_{system}}{V_{m,ref}}.
\end{equation}

\begin{equation}
\label{eq:NPPmin}
N_{PPmin} = \dfrac{N_{TPmin}}{N_{PSmin}}.
\end{equation}

\begin{equation}
\label{eq:NPS}
N_{PS} \geq N_{PSmin}.
\end{equation}

\begin{equation}
\label{eq:NPP}
N_{PP} \geq N_{PPmin}.
\end{equation}

\subsection{The Battery Storage Model }
\label{sec:BATmodel}

  
Various models have been described in the literature and the most common ones are based on lead-acid batteries~\cite{Copetti,Manwell93,Pinho};
that kind of battery has relative low cost and wide availability~\cite{Copetti}. 
Here, the model adopted uses only manufacturer's data without empirical tests~\cite{Copetti}. 
The discharge voltage equation is described by \eqref{eq:bat_Vd} as
%
\begin{multline}
\label{eq:bat_Vd}
V_{d} = \left[ 2.085-0.12(1-SOC) \right] - \\ \dfrac{I}{C_{10}} \left( \dfrac{4}{1+I^{1.3}} + \dfrac{0.27}{SOC^{1.5}}+0.02 \right) (1-0.007 \Delta T),
\end{multline}

\noindent where $C_{10}$ means 10h of rated capacity, which is standard on the manufacturer's data-sheet, $\Delta T$ is temperature variation ($\Delta T=T-T_{ref} $, $ T_{ref}=25^{o}C $), $ SOC $ or state of charge indicates how much electric charge is stored in the cell at a given time. Mathematically, it is the ratio between the present capacity and the nominal capacity (in $ Ah $, provided by manufacturer). If $SOC=1$, then the battery is totally charged; and if $ SOC=0 $, then the battery is fully discharged.  
%
%
The depth of discharge ($DOD$) or the fraction of discharge, is $DOC=1-SOC$.

%
%

For the charging process, however, the parameters are described by Eq.~\eqref{eq:Vcbat} as
\begin{multline}
\label{eq:Vcbat}
V_{c} = [2+0.16SOC]+ \\ \dfrac{I}{C_{10}} \left( \dfrac{6}{1+I^{0.86}} + \dfrac{0.48}{(1-SOC)^{1.2}} + 0.036  \right) (1-0.025 \Delta T).
\end{multline}

Note that SOC can be calculated easily at any point during the discharge period, thereby considering the current drained from batteries during a certain time period. 
In addition to the model verification, there is also the prior stage of project sizing check, as performed for the solar panel. Firstly we define the total capacity of the battery bank, as described by Eq. \eqref{eq:Cbank} as
\begin{equation}
\label{eq:Cbank}
C_{bank} = \dfrac{E_{corrected} \times autonomy}{V_{system} \times DOD}.
\end{equation}
\noindent where the variable $autonomy$ is a design definition and normally has a value ranging from $6$ to $48$h; the other variables were discussed previously in Section~\ref{sec:PVmodel} and ~\ref{sec:BATmodel}.
Secondly, the total (minimum) number of batteries is computed, as described by Eq.~\eqref{eq:Nbtotal}. Additionally, Eq.~\eqref{eq:batcheck} performs the final sizing check, thus considering the number of batteries in series ($ N_{BS} $) and the number of batteries in parallel ($ N_{BP} $) that are established to the project.
\begin{equation}
\label{eq:Nbtotal}
N_{B}total = N_{BS}min \times N_{BP}min = \dfrac{V_{system}}{V_{bat}} \times \dfrac{C_{bank}}{C_{20}}.
\end{equation}

\begin{equation}
\label{eq:batcheck}
\left( N_{BS} \times  N_{BP} \right) \geq N_{B}total.
\end{equation}

\subsection{Charge Controller Model}


%
%
%
%
In general, there are two main operating modes for the controller~\cite{Rawat}: normal operating condition, when the battery voltage fluctuates between maximum and minimum voltages; and overcharge or over-discharge conditions, which occur when the battery voltage reaches some critical values. 

To protect the battery against an excessive charge, the PV arrays are disconnected from the system, when the terminal voltage increases above a certain threshold $V_{max \textunderscore off}$ and when the current required by the load is less than the current delivered by the PV arrays~\cite{Hansen}. PV arrays are connected again when the terminal voltage decreases below a certain value $ V_{max \textunderscore on} $. 
%
%
In order to protect the battery against excessive discharge, the load is disconnected when the terminal voltage falls below a certain threshold $V_{min \textunderscore off}$ and when the current required by the load is larger than the current delivered by the PV arrays~\cite{Hansen}. The load is reconnected to the system, when the terminal voltage is above a certain value $V_{min \textunderscore on}$.
%
%
%
The steps in the modeling of the controller process are summarized in Table~\ref{table:controller}.

\begin{table}[!t]
\renewcommand{\arraystretch}{1.3}
\caption{Summary of the controller process (Source:~\cite{Hansen})}
\label{table:controller}
\centering
\begin{tabular}{c | c | c }
\hline
\hline
Step  & Constraint & Command\\
\hline
\hline
(1) & \makecell{If $V > V_{max \textunderscore off}$ \\and $I_{load} < I_{pv}$} & \makecell{Disconnect PV array \\from the system}\\
\hline
(2) & \makecell{If command (1) is \\done and $V < V_{max \textunderscore on}$} & \makecell{Reconnect PV array \\to the system}\\
\hline
(3) & \makecell{If $V < V_{min \textunderscore off}$ and \\ $I_{load} > I_{pv}$} & \makecell{Disconnect the load \\from the system}\\
\hline
(4) & \makecell{If command (3) is \\ done and $V > V_{min \textunderscore on}$} & \makecell{Reconnect the load \\to the system}\\
\hline
\hline
\end{tabular}
\end{table}

%
The output power ($ P_{out} $) of DC-DC converter is given by Eq.~\eqref{eq:poutcont} as
\begin{equation}
\label{eq:poutcont}
P_{in} = P_{out}.
\end{equation}

Assuming that the efficiency of the controller ($ \eta_{c} $) is a manufacturer's data, from Eq.~\eqref{eq:poutcont} we compute Eq.~\eqref{eq:potcont} as
\begin{equation}
\label{eq:potcont}
V_{in} I_{in} \eta_{c} = V_{out} I_{out},
\end{equation}

\noindent where $ V_{in} $ is the voltage across the PV array, $ I_{in} $ is the output current of PV array, $ V_{out}=V_{b}=V_{system} $ is the  DC bus voltage, and $ I_{out} $ is the output current from the converter.

%
%
%
%
%
%
%
 
One more time, some steps must be done to check the sizing project of the controller, prior the verification phase. Initially, the controller must meet the voltage requirement of the PV system, as described by Eq.~\eqref{eq:vcvsystem}: 
\begin{equation}
\label{eq:vcvsystem}
V_{c} = V_{system}.
\end{equation}

Following, the short circuit reference information from the manufacturer's solar panel must be corrected to the cell temperature, as described by Eq.~\eqref{eq:iscamb}:
\begin{equation}
\label{eq:iscamb}
I_{sc,amb} = I_{sc,ref} \times \left[ 1 + \eta_{I} \times (T-25) \right]. 
\end{equation}

The controller must meet the maximum current from the PV array given by \eqref{eq:icmin} and \eqref{eq:icicmin}.
\begin{equation}
\label{eq:icmin}
I_{c,min} = I_{sc,amb} \times N_{PP}.
\end{equation}

\begin{equation}
\label{eq:icicmin}
I_{c} \geq I_{c,min}.
\end{equation}

\subsection{The inverter model}
The role of the inverter is to keep the voltage constant on the AC side, i.e., at the rated voltage, 
and to convert the input power $ P_{in} $ into the output power $ P_{out} $ with the best possible efficiency $ \eta_{i} $ as described by Eq.~\eqref{eq:efficinv} \cite{Hansen}:
%
\begin{equation}
\label{eq:efficinv}
\eta_{i} = \dfrac{P_{out}}{P_{in}} = \dfrac{V_{AC} I_{AC} cos\varphi}{V_{DC}I_{DC}},
\end{equation}

\noindent where $ I_{DC} $ is the current required by the inverter from the DC source to be able to keep the rated voltage on the AC side, $ V_{DC} $ is the input voltage to the inverter delivered by the DC source (PV panel or battery),  $ V_{AC}  $ and $ I_{AC} $ are the output voltage and current, respectively, and $ cos \varphi $ can be obtained from the inverter's data-sheet.

%
The sizing project check of the inverter is carried out by means of three equations. Eq.~\eqref{eq:vindc} ensures that the input voltage of the controller meets the system voltage. Eq.~\eqref{eq:voutac} ensures that the output voltage of the controller meets the AC voltage of the load. Finally, Eq.~\eqref{eq:invcheck} ensures that the controller can support the total demand of the load and the surge power.
\begin{equation}
\label{eq:vindc} 
V_{in}DC = V_{system}.
\end{equation}
\begin{equation}
\label{eq:voutac} 
V_{out}AC = V_{AC}.
\end{equation}
\begin{equation}
\label{eq:invcheck} 
\left[ (Demand \leq P_{AC,ref}) \, and \, (P_{surge} \leq MAX_{AC,ref}) \right].
\end{equation}

\section{Automated Verification Using Model Checking}
\label{sec:AutomatedVerification}
Validation is the process of determining whether a design meets the user requirements, whereas verification is the process of determining whether a design meets a set of requirements, specifications, and regulations~\cite{ClarkeHV18}. If the requirements, specifications, and regulations are given in a formal language, then it may be possible to automate the verification process, thus resulting in a process known as \textit{formal verification}. Verification may form part of a validation process.
While simulation and testing explore some of the possible behaviors and scenarios of the system, leaving open the question of whether the unexplored trajectories may contain a flaw, formal verification conducts an exhaustive exploration of all possible behaviors. Thus, when a design is pronounced correct by a formal verification method, it implies that all behaviors have been explored, and the questions of adequate coverage or a missed behavior becomes irrelevant \cite{Clarke2012}.
 %

\begin{figure}[h]
\includegraphics[width=0.5\textwidth]{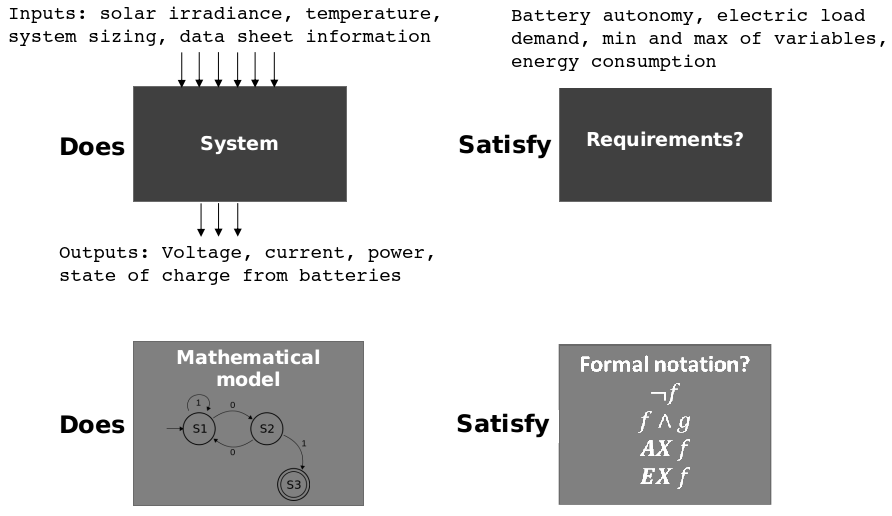}
\centering
\caption{From real system verification to model checking (adapted from \cite{ClarkeHV18}).}
\label{fig:systemverif}
\end{figure}

Formal verification is a systematic approach that applies mathematical reasoning to obtain guarantees about the correctness of a system \cite{Forejt2011}; one successful method in this domain is model checking \cite{Clarke2012}.
 %
%
The process of model checking can be split into three main components: modeling, specification, and verification method. In modeling, a model (normally mathematical) of the system is created; in specification, normally a list of properties to be satisfied by the system is established, i.e., the requirements, normally  expressed in a temporal logic form (e.g., CTL or LTL).
%
%
The model checking algorithm can be described as \cite{ClarkeHV18}:  
\begin{itemize}
\item Given the model $M$ and a CTL (or LTL) formula $ \phi $ as input;  
\item Model checking algorithm provides all the states of model $ M $ which satisfies $ \phi $;  
\item It returns \textit{YES} if $ \phi $ is \textit{TRUE}, or returns \textit{NO} if $ \phi $ is \textit{FALSE}.  
\end{itemize}
Specifically for the \textit{FALSE} verification result, the algorithm returns a \textit{counterexample} (i.e., a sequence of states that leads to a property violation), which is useful as diagnostic of the system to discover in which situation the model is violated; this is the most important advantage of the use of model checking~\cite{ClarkeHV18}. 
%
%
%
%
%
%
Fig.~\ref{fig:systemverif} shows the process to convert a real PV system to a model to be verified by a model checking procedure. 

However, there is a main disadvantage of model checking: the state explosion problem. In order to tackle this problem, many different techniques were developed in the last decades. One of the first major advances was symbolic model checking with binary decision diagrams (BDDs). In this approach, a set of states is represented by a BDD instead of by listing each state individually, which is often exponentially smaller in practice.
%
Another promising approach to overcome state explosion problem is Bounded Model Checking (BMC)~\cite{DBLP:conf/tacas/BiereCCZ99}. BMC is a method that checks a model up to a given path in the path length. BMC algorithms traverse a finite state machine for a fixed number of steps, $ k $, and checks whether a property violation occurs with this bound. It uses Boolean Satisfiability (SAT) or Satisfiability Module Theories (SMT) solvers to check the generated formula from BMC. 

SAT problem is a problem of determining whether there are certain conditions or interpretations that satisfy a given Boolean expression \cite{ClarkeHV18}. 
SMT decides the satisfiability of a fragment of first-order formulae using a combination of different background theories and thus generalizes SAT by supporting uninterpreted functions, linear and non-linear arithmetic, bit-vectors, tuples, arrays, and other decidable first-order theories~\cite{ClarkeHV18}.
The SAT or SMT solvers search the model for conditions (value of variables) that make the formula satisfiable. If a SAT or SMT solver finds a substitution for the formula/function then the substitute induces a counterexample and is said to be \textit{satisfiable}, i.e., it is satisfiable \textit{iff} the verified system contains errors.  
%
%
%
ESBMC is one of the most representatives bounded model checkers for embedded C/C++ software based on SMT solvers~\cite{esbmc2018}. 
ESBMC comes as an alternative to overcome limitations of the system modeling, especially considering that the system complexity is increasing and SMT has richer theories than SAT to represent models. 

\subsection{ESBMC}
ESBMC (or Efficient SMT-based Bounded Model Checker) is an open source, permissively licensed (Apache 2), cross platform bounded model checking for C and C++ programs~\cite{esbmc2018}, which supports the verification of LTL properties with bounded traces~\cite{DBLP:journals/sosym/MorseCN015}. 
ESBMC's verification flow can be summarized in three stages: (i) a front-end that can read and compile C/C++ code, where the formal specification of the system to be verified is first handled; (ii) preprocessing steps to deal with the representation of the code, control flow and unwinding of loops, and the model simplification, thereby aiming to reduce the verification effort; and finally (iii) the SMT solving stage, where all the constraints and properties of the system to be verified are encoded into SMT and checked for satisfiability.
If the SMT formula is shown to be satisfiable (SAT), a counterexample is presented; otherwise, the formula is unsatisfiable (UNSAT), i.e., there are no errors up to the given unwinding bound. 

ESBMC exploits the standardized input language of SMT solvers (SMT-LIB\footnote{http://smtlib.cs.uiowa.edu/} logic format) to make use of a resource called \textit{assertion stack}. An assertion, in SMT solvers, is a constraint over the variables in a formula that must hold if the formula is satisfiable~\cite{Morse2015}. New assertions can be added to or old assertions removed from this stack, depending on the evaluated value of variables. This enables ESBMC, and the respective solver, to learn from previous checks, optimizing the search procedure and potentially eliminating a large amount of formula state space to be searched, because it solves and disregards data during the process, incrementally. This technique is called ``incremental SMT''~\cite{DBLP:journals/fac/SchrammelKBMTB17} and allows us to reduce the memory overhead, mainly when the verified system is complex and the computing platform does not have large amount of memory to deal with all the design space state.

\section{Model Checking Stand-alone Solar Photovoltaic Systems }
\label{sec:Methodology}
\begin{figure}[h]
\includegraphics[width=0.4\textwidth]{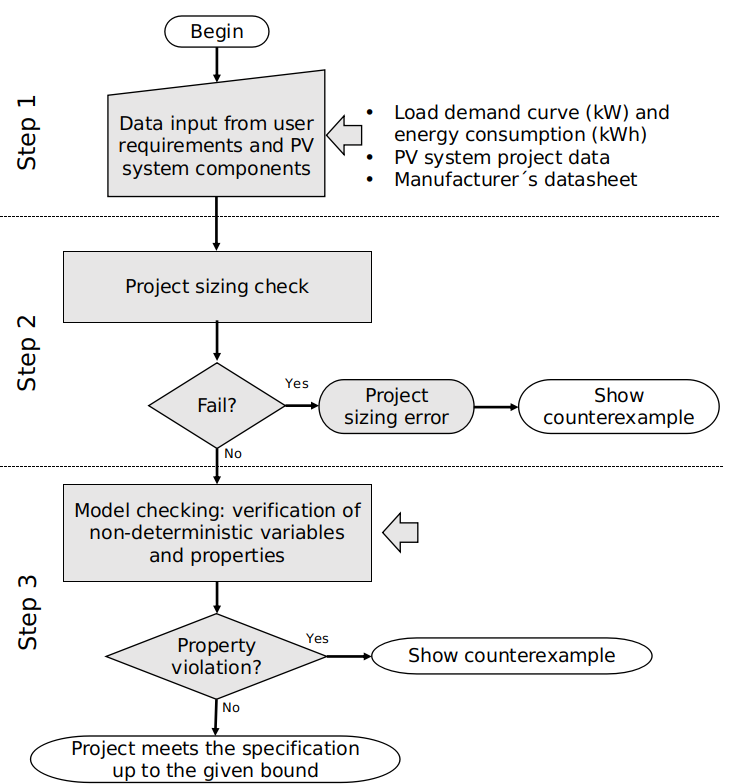}
\centering
\caption{Flowchart of the proposed automated verification of PV systems.}
\label{fig:flowchartgeneral}
\end{figure}

%
%
%
%
The flowchart of the proposed automated verification method is illustrated in Fig.~\ref{fig:flowchartgeneral}. 
In Step 1, the PV input data (e.g., load power demand and load energy consumption) and the formulae to check the sizing project, the mathematical model, the limits of the weather non-deterministic variables, are all written as an ANSI-C code~\cite{ANSI2018}. In Step 2, the sizing check of the PV system takes place to make sure that the components were selected according to the recognized design standards~\cite{Pinho}.
In Step 3, weather variables (e.g., solar irradiance and ambient temperature) will be systematically explored by our verification engine based on maximum and minimum values from the site, where the PV system will be deployed. 
In addition, depending on one of the desired properties of the system such as battery autonomy, energy availability, or even system power supply, our verification engine is able to indicate a failure if those properties are not met; in this particular case, it provides a diagnostic counterexample that shows in which conditions the property violation occurred. 

%
In a nutshell, ESBMC will process the ANSI-C code with constraints 
and properties 
from the PV system that are provided by the user, and the tool will automatically verify if the PV system requirements are met. If it returns a failure (i.e., SAT), then the tool provides a counterexample, i.e., a sequence of states that leads to the property violation; this information can be used as a feedback to improve the PV system design. However, if the verification succeeds (i.e., UNSAT), there is no failure up to the bound $k$; therefore, the PV system will present its intended behavior up to the bound $k$, i.e., our verification engine does not give any guarantee that there is no error in bound $k+1$ unless some induction method is employed~\cite{DBLP:journals/sttt/GadelhaIC17}.
%
%
%
%
%
%
%

Algorithm~\ref{alg:verification-algorithm} describes the pseudocode used to perform the automated verification. Line 1 indicates a function call that performs the size checking of the entire PV system: using Equations \eqref{eq:NTPmin}, \eqref{eq:NTP}, \eqref{eq:NPSmin}, \eqref{eq:NPS}, \eqref{eq:NPP}, and \eqref{eq:NPPmin} to verify the PV panel; using \eqref{eq:Cbank}, \eqref{eq:Nbtotal}, and \eqref{eq:batcheck} to verify the batteries; using \eqref{eq:vcvsystem}, \eqref{eq:icmin}, and \eqref{eq:icicmin} to verify the charge controller; and using \eqref{eq:vindc}, \eqref{eq:voutac}, and \eqref{eq:invcheck} to verify the inverter. The verification is carried out by the \textit{assert} macro from the ANSI-C programming language to encode each equation above. The argument to the \textit{assert} statement must be \textit{true} if the system specification is met; otherwise, the program aborts and prints a counterexample indicating a property violation. If there is no property violation, then the verification algorithm continues and the batteries are assumed to have SOC of 80\% (Line 5).

Information related to average temperature ($T$) and solar irradiance ($G$), maximum and minimum annual, are given to the algorithm in Lines 7 to 10 using non-deterministic variables from ESBMC to explore all possible states and the \textit{assume} macro to constrain the non-deterministic values using a given range. 
In order to reduce the computational effort of the algorithm,
every 24h-day was considered as a time-step of 1 hour, and it was split into two parts: (a) one where it is possible to occur PV generation, during daylight, with a duration in hours depending on each site (but dependent on the sun and weather conditions); and (b) one that includes all the remaining day (without any PV generation). Therefore, our approach depends on specific data about the solar irradiation levels to define the average amount of hours of PV generation.

After that, the model from PV generator is used in the function call of Line 11, to produce the voltage and current considering the states of $G$ and $T$. With respect to every hour considered, the conditional \textit{if-else-endif} statements from Lines 12, 17, 23 and 28, will perform the charge or discharge of batteries according to the value of different variables: if there is PV generation (which depends on $G$ and $T$), the updated state of charge from batteries, the house's load and the set-up information of the PV system.

Next, representing the time of the day where PV generation is not possible anymore, starting in Line 31, the algorithm will only discharge the batteries (using the 1 hour step) until a new charging process (at the next day) starts. Specific \textit{asserts} in Lines 27 and 35 will check the state of charging from batteries, and they will violate the property if their levels reach the minimum that represents a discharged battery; therefore, the PV system is unable to supply energy to the house. Nevertheless, if the verification engine does not fail, then we can conclude that the PV system does not need further corrections up to the given bounds.
%
 \begin{algorithm}
 \caption{Model checking algorithm for stand-alone PV}
 \begin{algorithmic}[1]
 \renewcommand{\algorithmicrequire}{\textbf{Input:}}
 \renewcommand{\algorithmicensure}{\textbf{Output:}}
  \STATE Perform project sizing verification( )
  \IF {(FAIL verification)} 
  \STATE exit (``Project sizing erro'')  
  \ENDIF
  \STATE $SOC \leftarrow 80\%$ \\
  \COMMENT {Starting with the PV generation time}
 \\ \textit{LOOP Process}
  \FOR {$h = 1$ to $Hours \, of \, PV \, generation$}
  \STATE $G \leftarrow nondet \textunderscore uint(\,)$ \COMMENT {$G$ is non-deterministic variable}
  \STATE $T \leftarrow nondet \textunderscore uint(\,)$ \COMMENT {$T$ is non-deterministic variable}
  \STATE assume ($Gmin \leq G \leq Gmax$) \COMMENT {restricting $G$ values}
  \STATE assume ($Tmin \leq T \leq Tmax$) \COMMENT {restricting $T$ values}
  \STATE $Imax, Vmax \leftarrow PVgenerationMODEL (G,T)$ 
  \\
  \COMMENT {Now, testing if battery is empty:}
  \IF {($SOC \leq SOC_{limit}$)} 
    \STATE assert (PV panel is generating energy?) \COMMENT {FAIL if not}
    \STATE $house \leftarrow energy \, from \, PV \, panels$
    \STATE $battery \leftarrow energy \, from \, PV \, panels$ 
    \STATE $SOC \leftarrow SOC + 1\,h\, charge$
  \ELSIF {($PV \, Array  \geq Vbulk$)}
  	\STATE depending on $SOC$, Battery enter in absorp. or float.
	\STATE adequate the voltage at DC-bus (PV panel feed bus)
    \STATE $house \leftarrow energy \, from \, PV \, panels$
    \STATE $battery \leftarrow energy \, from \, PV \, panels$ 
    \STATE $SOC \leftarrow SOC + 1\,h\, charge$
  \ELSE
  	\STATE Adequate the voltage of DC-bus (battery feed bus)
	\STATE $house \leftarrow energy \, from \, batteries$
    \STATE $SOC \leftarrow SOC - 1\,h\, discharge$
    \STATE assert ($SOC \geq SOC_{limit}$)
    \\
    \COMMENT {this ELSE: batteries $\geq SOC_{limit}$ but panels are off}	  
  \ENDIF
  \STATE $h \leftarrow (h+1)$
  \ENDFOR
 \\ \textit{Start of battery autonomy verification:}
\STATE $AutonomyCount \leftarrow 1$
 \WHILE {$AutonomyCount \leq autonomy$}

  \STATE $SOC \leftarrow SOC - ( 24 - Hours \, of \, PV \, generation)\,h\, discharge$
  \STATE $AutonomyCount \leftarrow ( 24 - Hours \, of \, PV \, generation)$
  \\  
    \COMMENT {autonomy verification during discharge period}
  \\
  \STATE assert ($SOC \geq SOC_{limit}$)  
  \\
  \COMMENT {Perform similar \textbf{for}-LOOP as defined in line 6}
  \ENDWHILE
 \RETURN $(\,)$ 
 \end{algorithmic} 
 \label{alg:verification-algorithm}
 \end{algorithm}
\section{Experimental Evaluation} 
\label{sec:results}
%
%
%

\subsection{Description of the Case Studies}

We have performed five case studies to evaluate our proposed verification method: (a) four PV systems (700 W inverter, with 48h autonomy) deployed in four different houses in an indigenous community (GPS coordinates 2$^{o}$44'50.0"S 60$^{o}$25'47.8"W) situated nearby Manaus (Brazil), with each house having a different power demand (house 1 = 253 W, house 2 = 263 W, house 3 = 283 W, and house 4 = 501 W); and (b) one case concerning a system deployed as an individual system in Manaus (GPS coordinates 3$^{o}$4'20.208"S 60$^{o}$0'30.168"W), supporting 915 W of the house's load (house 5 with 1,200 W inverter, and autonomy of just 6 h). 

Note that the annual average temperature ($T$) in Manaus is from 23$^{o}$C to 32$^{o}$C; and irradiance ($G$) varies from 274 W/m$^{2}$ to 852 W/m$^{2}$ when there is sunlight (that information is provided in Lines 9 and 10 of Algorithm~\ref{alg:verification-algorithm}). Another characteristic of Manaus, based on historical weather data \cite{Temperature}, \cite{Irradiance}, is related to the fact that only during 8 hours of the day is possible to have PV generation, from 8:00h to 16:00h (that information is provided in Algorithm~\ref{alg:verification-algorithm} as well).

\subsection{Objectives and Setup}
Our experimental evaluation aims to answer two research questions:
\begin{enumerate}
\item[RQ1] \textbf{(soundness)} Does our approach provide correct results?
\item[RQ2] \textbf{(performance)} How does our approach compare against other existing tools?
\end{enumerate}

In order to evaluate the proposed verification method and its performance, we have considered five case studies and also compared its performance to the HOMER Pro tool. Every dweller, who owns a PV system, was interviewed to get information about his/her PV system during four months of use. This information was used to know possible flaws from every system in the field.

All experiments were conducted on an otherwise idle Intel Core i7-2600 (8-cores), with 3.4 GHz and 64 GB of RAM, running Ubuntu 18.04.1 LTS 64-bits. Concerning our verification engine, ESBMC v5.1 was used with the SMT incremental mode\footnote{The command-line used to perform the verification is: \$ esbmc filename.c -\phantom{}-no-bounds-check -\phantom{}-no-pointer-check -\phantom{}-no-div-by-zero-check -\phantom{}-unwind 300 -\phantom{}-smt-during-symex -\phantom{}-smt-symex-guard -\phantom{}-z3} enabled with the goal of reducing memory usage; we have also used the SMT solver Z3 version 4.7.1~\cite{DeMoura}. The experiments were performed without a predefined timeout.

Experimental setup of HOMER Pro: all experiments were conducted on an otherwise idle Intel Core i5-4210 (4-cores), with 1.7 GHz and 4 GB of RAM, running Microsoft Windows 10; we have used HOMER Pro v3.12.0.

\subsection{Results}
\label{sec:results_indeed}
%
%
The description of our experimental results can be broken down into two parts: (a) the 1,200 W PV system (house 5) failed during the sizing check since the number of panels was \textit{incorrectly} sized; in particular, the counterexample provided by our verification engine indicated 3 PV panels in parallel and the actual project has 2 in series and 2 in parallel. This verification took approximately 63.3 hours to be performed. Surveying the owner of the 1,200 W system we identified that, in fact, the system does not meet the battery autonomy most of the time (mainly when all loads are turned on), thus affirming RQ1; this behavior is expected since the system was purchased as an off-the-shelf solution and not as a customized design for the electrical charges of the house; (b) For the 700 W PV systems of houses 1, 2, 3, and 4, the sizing check was successful during verification, but our verification engine has found flaws related to the battery autonomy, particularly when SOC reached a empty-battery level. Our verification engine identified those flaws (for all four houses) right after the first night-discharge cycle, i.e., before the solar system started to recharge the batteries. Our verification engine took approximately 409.3 hours to find this flaw in house 1; 611.2 hours for house 2; 615.8 hours for house 3, and 620.8 hours for house 4. These flaws were confirmed with the dwellers who own the systems by an interview process: at least once a month is usual the system to turn off, normally in raining or clouds days, thereby reaching the situation described in Step 3 of Table~\ref{table:controller}, further affirming RQ1; after the sun rises, the systems returns to normal condition operation. 

The same five case studies were evaluated by HOMER Pro. The simulation results showed that the project restrictions were met by four 700 W PV systems (house 1, 2, 3 and 4), without any indication of sizing error or even performance related issues. The case study that was unsuccessful during simulation was the 1,200 W (house 5); however, without any indication about the failures of this PV system. All the simulations took less than 5 seconds (each) to be performed by HOMER Pro.
Despite the divergence of results for the houses 1, 2, 3 and 4 w.r.t. our proposed approach, it is evident that the information collected from the dwellers indicate that our approach provides the correct evaluation of the PV system, thus answering RQ2. House 5 presented flaws from both tools; however, only our approach indicated which design error was responsible for the flaw (number of PV panels), further answering RQ2.
Note that a PV design always uses daily average values of sun hours to each site, with impact in the PV components. Those hours are based on historical data and, in field, it is not unusual to find days where that number of hours was not reached due to weather conditions. The season has impact since the case studies are from the rain forest, where clouds are always present. As a result, the identified flaws in houses 1, 2, 3, and 4, are justified once again.
\subsection{Threats to Validity}
We have reported a favorable assessment of the proposed method over a diverse set of real-world benchmarks. Nevertheless, we have also identified three threats to the validity of our results that can further be assessed.

\textit{Model precision:} each component of the PV system is mathematically modelled, and the precision of the proposed method depends on the precision of that particular model. A careful evaluation in a PV laboratory to validate the model could add more reliability to the results produced by our method.

\textit{Time step:} The run-time complexity of our proposed method is an issue; the time step of one hour can be further reduced to approximate the algorithm to the real-world scenario, where a solar irradiance and ambient temperature can change in fractions of minutes.

\textit{Case studies:} Our case studies are performed only in Manaus, in particular in the south hemisphere. A more complete evaluation can be performed if other places around the world could become case studies.

\section{Conclusion and Future Work}
\label{sec:Conclusions}

We have described and evaluated an automated verification method to check whether a given PV system meets its specification using software model checking techniques. We have considered five case studies from real photovoltaic systems deployed in five different sites, ranging from $700$ W to $1,200$ W. Although this verification method takes longer than simulation methods, it is able to find specific conditions that lead to failures in a PV system previously validated by a commercial simulation tool. In particular, the proposed method was successful in finding sizing errors and indicating in which conditions a PV system can fail. 
As future work, the proposed method will be extended to start from a list of commercial equipment, where each equipment is verified and the final solution, which satisfies the project specification, is found via Counterexample Guided-Inductive Synthesis~\cite{DBLP:conf/asplos/Solar-LezamaTBSS06}, thus leading to an optimum sizing of PV systems. We will also consider other types of renewable energy and even hybrid ones to allow our method to design and verify typical rural electrification.
%



%



\section*{Acknowledgment}
To Coventry University and Sustainable Amazonas Foundation (FAS) for the possibility to test real PV systems.

\begin{IEEEbiography}
    [{\includegraphics[width=1in,height=1.25in,clip,keepaspectratio]{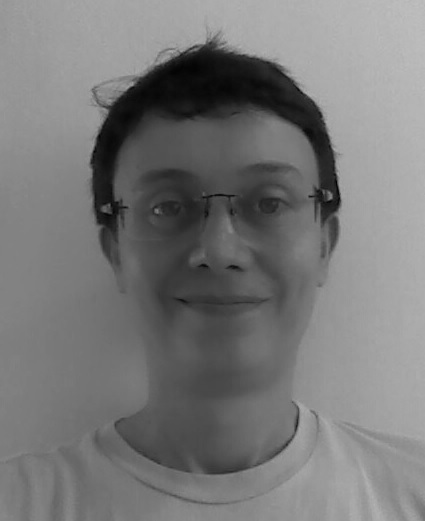}}]{Alessandro Trindade}
received his BSc and MSc in Electrical Engineering from the Federal University of Amazonas (UFAM) in 1995 and 2015, respectively. Currently, he is pursuing his PhD in the Postgraduate Program in Informatics (PPGI) at UFAM, and holds an Assistant Professor position in the Electricity Department from UFAM. Prior to joining UFAM, he worked 4 years as Consultant of renewable energy to the State Electric Utility and to the Inter-American Institute for Cooperation on Agriculture (IICA); he also worked for 12 years as R\&D and project manager at a non-profit foundation (Centre of Analysis, Research and Innovation Technology Foundation). His interest is in renewable energy, automated verification, and model checking.
\end{IEEEbiography}
%
%
\begin{IEEEbiography}
    [{\includegraphics[width=1in,height=1.25in,clip,keepaspectratio]{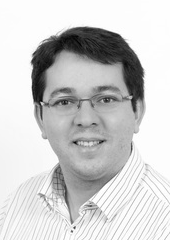}}]{Lucas Cordeiro}
received his Ph.D. degree in Computer Science in 2011 from the University of Southampton, UK. Currently, he is a Senior Lecturer in the School of Computer Science at the University of Manchester, UK and leads the Systems and Software Verification laboratory. He is also a collaborator in the Postgraduate Program in Electrical Engineering and Informatics at the Federal University of Amazonas (UFAM), Brazil. Prior to joining the University of Manchester, he worked as a researcher / researcher engineer at Oxford University / Diffblue and as an adjunct professor at UFAM; he also worked for 4 years as a software engineer at Siemens Mobile and CT-PIM. His work focuses on software model checking, automated testing, program synthesis, and embedded \& cyber-physical systems.
\end{IEEEbiography}






\end{document}